# Measurement of 3D bubble distribution using digital inline holography

Siyao Shao[1, 2], Jiarong Hong[1, 2, *]

1. Saint Anthony Falls Laboratory, 2 3rd AVE SE, University of Minnesota, Minneapolis, MN, USA 55414.

2. Department of Mechanical Engineering, 111 Church ST SE, University of Minnesota, Minneapolis, MN, USA 55414.

* Email addresses of the corresponding author: jhong@umn.edu

## Abstract

The paper presents a hybrid bubble hologram processing approach for measuring the size and 3D distribution of bubbles over a wide range of size and shape. The proposed method consists of five major steps, including image enhancement, digital reconstruction, small bubble segmentation, large bubble/cluster segmentation, and post-processing. Two different segmentation approaches are proposed to extract the size and the location of bubbles in different size ranges from the 3D reconstructed optical field. Specifically, a small bubble is segmented based on the presence of the prominent intensity minimum in its longitudinal intensity profile, and its depth is determined by the location of the minimum. In contrast, a large bubble/cluster is segmented using a modified watershed segmentation algorithm and its depth is measured through a wavelet-based focus metric. Our processing approach also determines the inclination angle of a large bubble with respect to the hologram recording plane based on the depth variation along its edge on the plane. The accuracy of our processing approach on the measurements of bubble size, location and inclination is assessed using the synthetic bubble holograms and a 3D printed physical target. The holographic measurement technique is further implemented to capture the fluctuation of instantaneous gas leakage rate from a ventilated supercavity generated in a water tunnel experiment. Overall, our paper introduces a low cost, compact and high resolution bubble measurement technique that can be used for characterizing low void fraction bubbly flow in a broad range of applications.

## 1. Introduction

Bubbly flow occurs in a wide range of natural settings (e.g., deep-sea venting) as well as industrial applications (e.g. petroleum refining process, undersea natural gas exploitation and air-lift reactors in bio-chemical processes). The characteristics of bubbles, including its size distribution and morphology, strongly influence the fundamental physical processes in such flows. Specifically, for example, in a horizontal pipe turbulent bubbly flow, the bubble size distribution and morphology determine the pressure drop and wall heat transfer (Kamp et al., 2001). Kamp et al. (2001) explored the influences of bubble size distribution on bubble induced turbulence and vice-versa in pipe flow. Subsequently, Liao et al. (2009 and 2010) summarized the existing models relating bubble size and the turbulence in the flow and its influence on the bubble coalescence and breakup. For the various applications involving multiphase chemical reaction and aerations, bubble size distribution is the key factor that influences the chemical reaction rate and mass transfer processes through affecting interfacial area of the bubbles (Roseler and Lefebvre 1989, Smith et al. 1996, Junker 2006, and Lau et al. 2013). Particularly, a recent investigation of hydroturbine aeration modeled the gas mass transfer rate based on bubble size distribution, which correctly predicted the varying saturation of the oxygen in the water over time due to aeration (Karn et al. 2015a). Therefore, an accurate measurement of the distribution of bubble size and morphology is crucial for both the fundamental study of the characteristics of bubbly flows and various associated industrial applications.

The techniques for bubble measurements include both intrusive and non-intrusive approaches. Up to date, industrial applications rely largely on the intrusive approaches that measure individual bubbles or void fraction of bubbles in a bubbly flow though they could cause substantial disturbance on the flow field and bubble distribution (Felder and Chanson 2015, Freddy et al. 2018). For individual bubble measurement, optical probe method employs an optic fiber which emits light and receives reflected signal to detect the presence of bubbles in the flow, and uses the correlation of the signals from two spatially-separated probes to determine bubble velocity and size (Saberi et al. 1995). Similarly, the capillary suction probe uses a photodetector to detect the bubble slugs introduced by a vacuum into the capillary tube which affects the local pressure in the flow (Laakkonen et al. 2005). The mapping of void fraction of the bubbly flow can be achieved through the measurement of electrical or capacitance signals across different ends of a wire mesh inserted in the flow (Liu and Bankoff 1993, Da Silva et al., 2007).

In general, non-intrusive techniques, particularly imaging based methods are preferred over the intrusive ones although they are usually limited to relatively low void fraction of bubbles. In the literature, bubble images are typically recorded through shadowgraph method, which employs a volumetric backlighting to generate shadow images of bubbles on the camera sensor (Estevadeordal and Goss 2005). Bubble shadowgraph often faces challenges such as non-uniformity in background lighting, the presence of dense bubble clusters, and the coexistence of in-focus and out-of-focus bubbles over a wide range of size. In addition, it is essentially a 2D imaging technique with limited depth-of-field (DOF). For bubble shadowgraph, the bubble size distribution is usually obtained through image segmentation with a global or local intensity thresholding (e.g., Otsu 1979, Sahoo et al. 1997, Honkanen et al. 2005, and Lau et al. 2013). Then, the segmentation of the bubble clusters can be achieved by approximating the overlapping bubbles through an object recognition approach which fits ellipsoidal shapes to the object areas (Honkanen et al. 2005). Some other reports focused on implementing watershed algorithm to distinguish individual bubbles from bubble clusters (Bonifazi et al. 1999, Lin et al. 2008, and Lau et al. 2013). However, such approach is observed to have over-segmentation issues which can be mitigated using marker-controlled segmentation methods (Gonzales et al. 2009). To tackle the issue of segmenting in-focus and out-of-focus bubbles over a large size range, Karn et al. (2015b) introduced a hybrid method that first categorize the bubbles based on their size, and employed different segmentation approaches for bubbles within each size range. Particularly, a new cluster segmentation process was developed which combines watershed segmentation and multiple morphological operators to extract both in-focus and out-of-focus individual bubbles from bubble clusters. It is worth noting that the bubble image can also be obtained using laser sheet illumination. Nevertheless, the image quality resulting from such method can be substantially deteriorated by the strong light refraction and reflection at bubble surfaces (Brücker 2000, Wosnik and Arndt 2013), which is not commonly used for imaging high concentration bubbly flows.

Recently, digital inline holography (DIH) was introduced as a low cost and compact solution for measuring bubble size distribution in 3D (Tian et al. 2011, Tapapatra et al. 2012, Liu et al. 2013, Sentis et al. 2018). DIH employs a single beam light source to illuminate the objects and uses a digital sensor to record the holograms generated from the interference between the scattered light from the objects and un-scattered portions of the beam. The 3D information of the objects is obtained through digital reconstruction and object segmentation (Katz and Sheng 2010). Various approaches have been proposed to segment bubbles from the reconstructed holograms.

Specifically, for example, Tian et al. (2011) employed minimum intensity metric on the bubble edge in the holograms to determine the bubble depth and Gaussian mixture model to discriminate individual bubbles from their clusters. The hologram processing program was assessed through measuring a well-mixed bubbly flow with mostly spherical bubbles in a static water tank. Similar approach was also implemented by Sentis et al. (2018) for measuring a mixture of diluted bubbles and oil droplets rising in a quiescent water tank, capable of distinguishing the signature of bubbles and oil droplets based on their minimum intensity metric. In addition, multiple investigations have attempted to incorporate additional segmentation criteria such as shapes or intensity gradients to improve the accuracy of segmentation and localization of the bubbles from holograms (e.g. Tapapatra et al. 2012, Liu et al. 2013). Particularly, Tapapatra et al. (2012) combined intensity metric and intensity gradient with circular Hough transform to segment bubbles from the holograms. The method assumes spherical bubbles in the flow and determines the depth bubble based on pixel intensity gradient calculated from a Sobel filter. Using this approach, it has conducted the measurement of the bubble size distribution in a ship wake using holograms captured by an undersea DIH setup. Nevertheless, according to Zhang et al. (2006) and our preliminary tests on our water tunnel bubble holograms with a wide range of bubble sizes and shapes, the robustness of previous approaches drops substantially with increasing complexity of bubble field and background noises. Additionally, none of the existing DIH-based bubble measurement approaches can characterize the orientation of bubbles within the holograms which is an important parameter for estimating the gas volume contained in large irregular bubbles. Khanam et al. (2011) introduced a method to determine the longitudinal tilting for needle-shaped particles through DIH. However, such method is applied for measuring small particles with well-defined fringe patterns in the hologram and has not been assessed with bubble measurements.

In the present paper, we introduce a new DIH bubble segmentation approach to address the abovementioned issues. The approach is implemented to analyze the holograms of bubbly wakes obtained from a ventilated supercavitation experiment, which contain bubbles over a high dynamic range of size and shape as well as bubble clusters. The current paper is structured as follows: Section 2 provides the experimental facility and the DIH measurement setup for generating the bubble holograms. In Section 3, the proposed DIH processing algorithm for extracting the 3D location and size distribution of the bubbles from the holograms is described. Subsequently, the assessment of the proposed algorithm using synthetic bubble holograms and a 3D printed physical target is provided in Section 4, which is followed by a demonstration of the proposed approach using the data from water tunnel bubbly wake measurements in Section 5. Additionally, Section 6 is a discussion and conclusion of the presented results.

## 2. Experimental setup

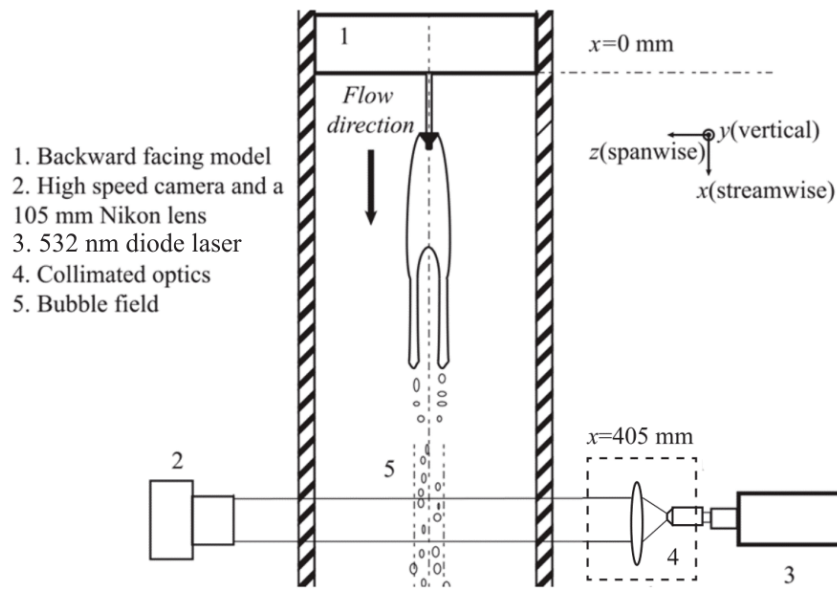

Fig.1. Schematic showing the measurements of bubble distribution in a bubbly wake generated from ventilated supercavitation using a DIH setup.

The bubble holograms used in the current study is produced in a ventilated supercavitation experiments as shown in Fig. 1. The experiments are conducted in a recirculating cavitation water tunnel at the Saint Anthony Falls Laboratory (SAFL). This tunnel has been used for a number of supercavitation and hydrofoil aeration experiments in recent years (Karn et al. 2015a, Karn et al. 2016a, and Shao et al. 2018). The detailed specifications of the water tunnel can be referred to the Shao et al. (2018). A backward facing model with 10 mm-in-diameter cavitator are introduced to generate the cavity. The velocity of the water flow is set at 4.0 m/s with a maximum uncertainty of 0.1 m/s (Karn et al. 2016a). It is worth noting that in the current investigation, we use a mass flow controller to regulate the input ventilation with the unit of liter per minute under standard condition (SLPM) where standard condition corresponds to a temperature of 273 K and a pressure of 101.3 kPa (Karn et al. 2016a). The details of the backward facing model and the flow conditions can be referred to Karn et al. (2016a).

The high speed DIH setup comprises a 532 nm continuous diode laser, beam expansion optics and an APS-RX high speed camera with a Nikon 105 mm imaging lens (Fig. 1). Particularly, the beam expansion optics includes a 20× magnification objective lens for focusing and diverging the beam from the laser diode and a convex lens of 45 mm focal length to collimate the laser into a 50 mm-in-diameter beam. To capture the bubbles moving with a relatively high speed (~ 4 m/s), the high-speed camera is set to be operated under 3000 fps with a shutter speed 9.3 µs to generate clear fringe patterns with the bubbles and free of blurring. The magnification of the lens is set to be 1:3 corresponding to a pixel dimension of 51 µm/pixel. The DIH system is placed 405 mm downstream of the trailing edge of the supporting hydrofoil of the backward facing model. Over 8000 holograms are captured through the span of the experiments corresponding to a time scale of 2.71

s. The bubble sizes are in a range of 2 to 180 pixels, which correspond to a bubble diameter of 0.1 mm to 9 mm.

## 3. Proposed Methods

### 3.1. Overview of the hologram processing task

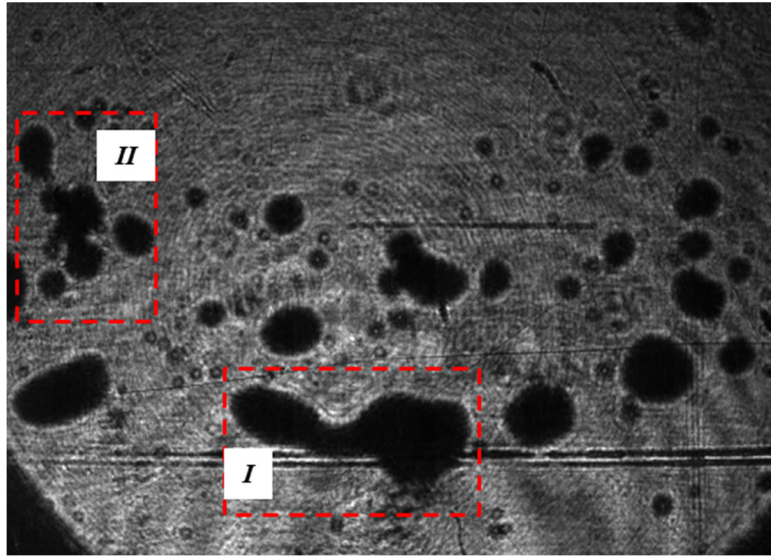

Fig.2. A sample of original bubbly wake hologram from our experiments. Note the irregular (dumbbell shape) bubble in dashed box *I* and a bubble cluster in dashed box *II*.

Fig. 2 presents a sample of the original hologram in the experiments to demonstrate a number of challenges in the processing of holograms from real water tunnel experiments, which has not been adequately addressed by the previous techniques yet (Tian et al. 2011, Tapapatra et al. 2012, Liu et al. 2013, Sentis et al. 2018). These challenges include strong background noise, high dynamic range of the bubble size and shape, and bubble clustering as shown in the supplemental video S1. Specifically, the background noises include temporal fluctuation of background intensity and interference noises caused by scratches on the tunnel window. In particular, the temporal fluctuation of background intensity is attributed to the fluctuating intensity of laser illumination, and more significantly, the temporal varying bubble void fraction in the field of view. Further, the holograms contain irregular shape bubbles (i.e., dumbbell shape and elliptical) with a broader size range (corresponding to equivalent bubble diameter from 100 μm to 10 mm) in comparison with the previous investigations. The bubble clusters appeared in the field view include both the bubbles that are physically clustered together and those generated from the cross-interference of the neighboring bubbles. To address all these challenges, a new hybrid hologram processing approach is developed as detailed in the section below.

## 3.2. Hybrid hologram processing approach

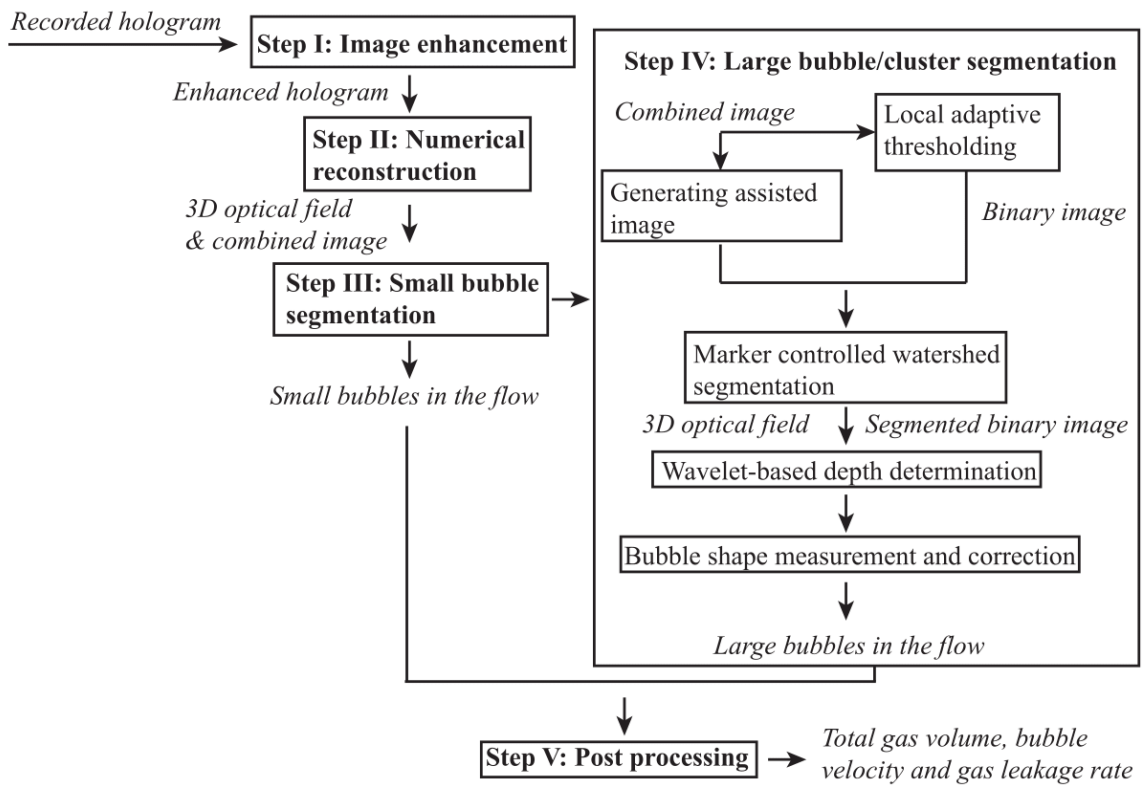

Fig. 3. Schematic showing the steps involved in the proposed hybrid hologram processing approach.

Fig. 3 shows the general steps involved in the proposed hybrid hologram processing approach. The proposed approach consists of five major steps: (i). image enhancement; (ii). numerical reconstruction; (iii). small bubble segmentation; (iv). large bubble/cluster segmentation; (v). post processing.

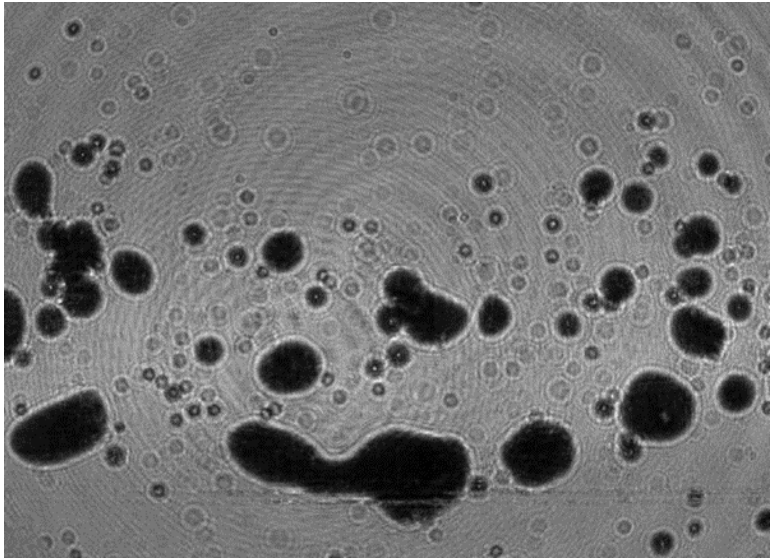

Fig. 4. The enhanced bubble hologram after background removal.

As the first step, hologram is enhanced by subtracting the background of the holograms (i.e. the stationary part of the hologram) from each individual one followed by an intensity normalization afterwards. The background of the hologram is obtained through an ensemble average of the holograms in a sequence with a skip determined by the average flow velocity in streamwise direction (i.e. use holograms every other 18 frames in the current case). Fig. 4 presents the result of the hologram (Fig. 2) after enhancement. As it shows, the stationary noise existing in the hologram is removed from each individual hologram. It is worth noting that the enhanced hologram shows a global fringe pattern (i.e., concentric circles in the background of Fig. 4) due to the fluctuation of light source energy profile which will not affect the processing of the hologram.

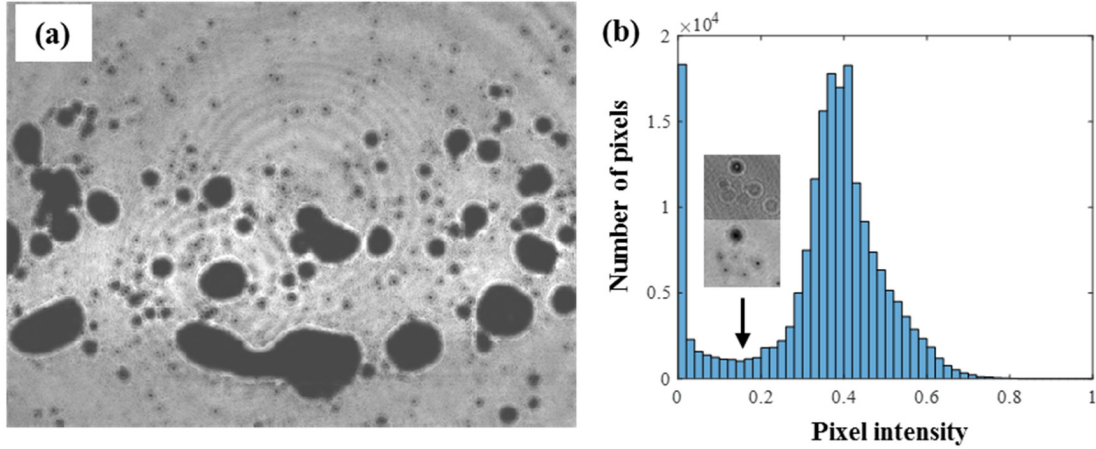

Fig. 5. (a) Combined image after digital reconstruction of Fig. 4 and (b) its pixel intensity histogram. Note that some small bubbles have pixel intensity located between the two peaks of the bi-modal intensity histogram. The inset figures in (b) show the original hologram sample of small bubbles cropped out from Fig. 4 and the corresponding combined image after reconstruction cropped out from Fig. 5(a)

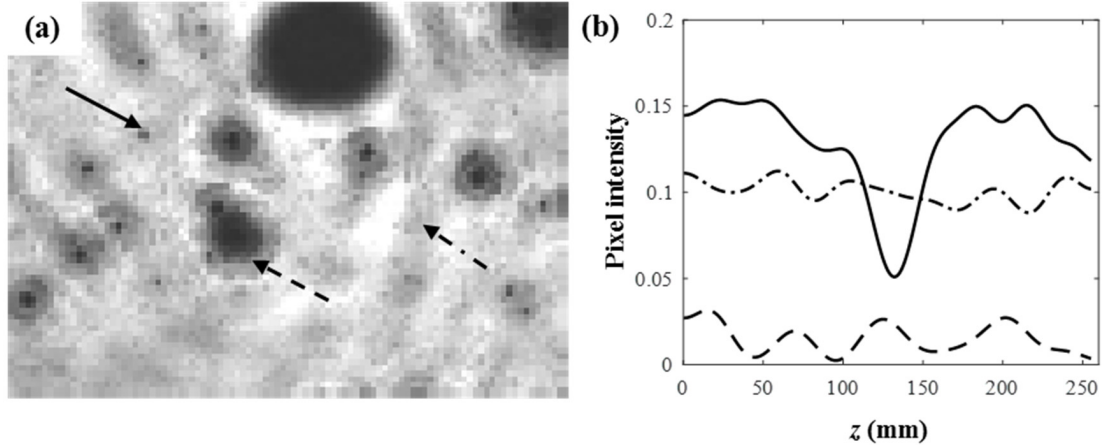

Fig. 6. (a) A subregion of the Fig. 5(a) depicts three types of pixels corresponding to small spherical bubbles (solid arrow), large bubbles/bubble clusters (dashed arrow) and background (dash-dot arrow). (b) Typical longitudinal intensity profile of a pixel corresponding to small spherical bubbles (solid curve), large bubbles/bubble clusters (dashed curve) and background (dot-dashed curve).

Subsequently, the enhanced hologram is numerically reconstructed. This step generates a 3D reconstructed optical field and its corresponding 2D combined image (i.e., longitudinal minimum intensity projection of each pixels in *xy* plane) from the original hologram for processing in the later steps. The reconstruction of the hologram to 3D optical field uses Rayleigh-Sommerfield formula:

$$u_p(x, y, z) = I_k(x, y) \otimes h(x, y, z) \tag{1}$$

where $u_p$ is the 3-D complex optical field calculated through reconstruction and $x, y, z$ are lateral and longitudinal directions, respectively. In addition, the $I_k$ refers to the enhanced hologram, $\otimes$ is the convolution operator and $h$ the Rayleigh-Sommerfield Kernel with setting the cosine term to unity for simplification (Katz and Sheng 2010):

$$h(x,y,z) = \frac{1}{j\lambda\sqrt{x^2+y^2+z^2}} \exp[jk(\sqrt{x^2+y^2+z^2})] \qquad (2)$$

The $\lambda$ in the above equation is the wavelength of the illumination beam and $k$ is the wave number. The convolution integral is usually calculated as a simple multiplication in Fourier domain using fast Fourier transform as below:

$$u_p(x,y,z) = FFT^{-1}\{FFT[I_k(x,y)] \times FFT[h(x,y,z)]\} \qquad (3)$$

where $FFT[\,]$ represents the fast Fourier transform operator. The complex optical field $u_p(x,y,z)$ in the above equation is the 3D image of the objects. Corresponding pixel intensity at each location can be calculated through: $I(x,y,z) = u_p(x,y,z) \times conj[u_p(x,y,z)]$, where $conj[\,]$ represents the conjugate operator. The combined image after reconstruction is then generated through the projection of minimum intensity in the longitudinal direction of individual pixels onto $xy$-plane (Fig. 5a). As illustrated by Fig. 5(b), after reconstruction, the pixel intensity distribution in the combined image shows a peak near zero which is corresponding to the bubbles and a peak located rightwards refers to background pixels. Additionally, for some small spherical bubbles, the intensity of their pixels in the combined image is located between these two peaks. As shown in Fig. 6, a scanning of pixel intensity value in longitudinal direction demonstrates that the pixels corresponding to small bubbles exhibit a prominent local minimum at its in-focus longitudinal location with a value much lower than the background pixels (solid line). Comparatively, the pixel intensity of neither large bubbles nor background shows prominent minima along the longitudinal direction. Such difference in the pixel intensity variation in the longitudinal direction allows us to separate bubbles into two groups based on their size, i.e. small bubbles and large/cluster bubbles for the following processing to determine their location and size, etc.

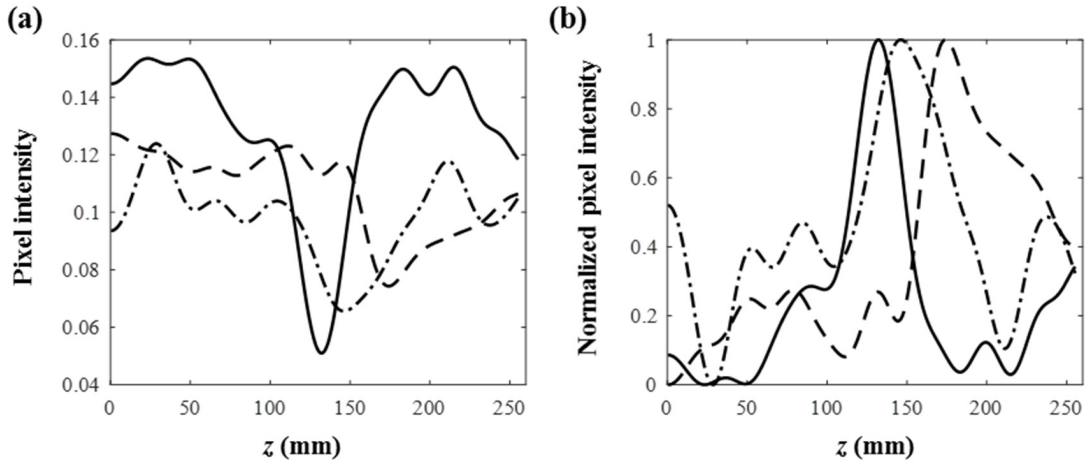

Fig. 7. The longitudinal intensity profile of a sample pixel that belongs to a small bubble (a) before and (b) after intensity inversion and normalization.

Based on the above discussion, we extract the small bubbles from the holograms by identifying the connected regions of pixels that show a prominent minimum in their longitudinal intensity profile, and the *z*-location of each bubble is determined by averaging the location of the prominent minimum of each pixel within the bubble. Specifically, to determine the pixels with prominent minimum, the intensity of each pixel along the longitudinal direction is first inverted and then normalized to enhance the signal-to-noise ratio (SNR) (Fig. 7). Based on Grubbs (1969), the highest peak of the intensity profile with an intensity 30% higher than the second highest peak is considered as a prominent peak corresponding to the prominent minimum and also the in-focused position of the pixel. Each individual small bubble is then identified through labelling connected pixels that show prominent minimum in their longitudinal intensity profile. The resulted small bubble field is stored as a 2D binary image. The size and 2D centroid locations ($x$, $y$) of each small bubble are calculated based on the area occupied by their pixels ($A$), and its diameter ($d$) is obtained by assuming the bubble to be spherical using $d = \sqrt{4A/\pi}$.

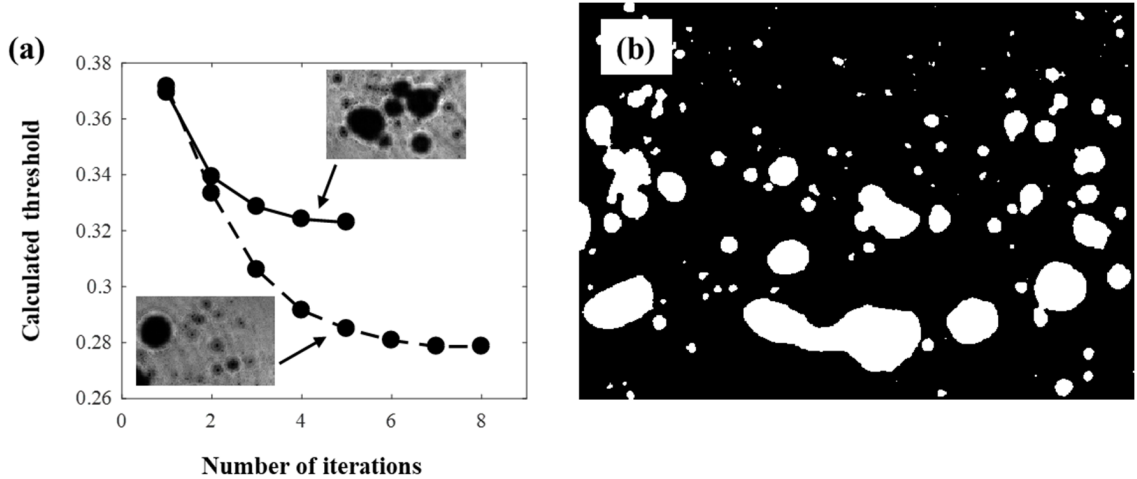

Fig. 8. (a) Iterations of the threshold values for the two subregions from Fig.5 (a) and the corresponding thresholded result of large bubbles/clusters.

The large bubbles/clusters are extracted from a hologram by first binarizing the combined image with a local adaptive thresholding method and excluding pixels of the small bubbles (extracted from the abovementioned approach) from the binarized image. Specifically, this local adaptive thresholding method employs a sliding window with size sufficient to encompass the largest bubble in the hologram sequence, and detemine the binarization threshold for the center pixel in each window based on the pixel intensity distribution within the window. The optimal threshold value for each window is calculated through an iterative process following Gonzalez et al. (2009). First, it initiates the thresholding using the mean pixel intensity inside the window, which divides the pixels in the window into two groups. Then it computes the mean pixel intensity for each group and determines the new threshold to be the average of the two mean intensities. The process then repeats until the threshold converges as shown in Fig. 8(a). For pixels near the borders of the image, reflective padding (i.e. filling the pixel values beyond the region of the image by reflecting pixel values of the image with respect to each border line) is used to ensure a proper thresholding. The Fig. 8(b) presents a binarized sample image corresponding to Fig. 5(a) with small bubbles excluded.

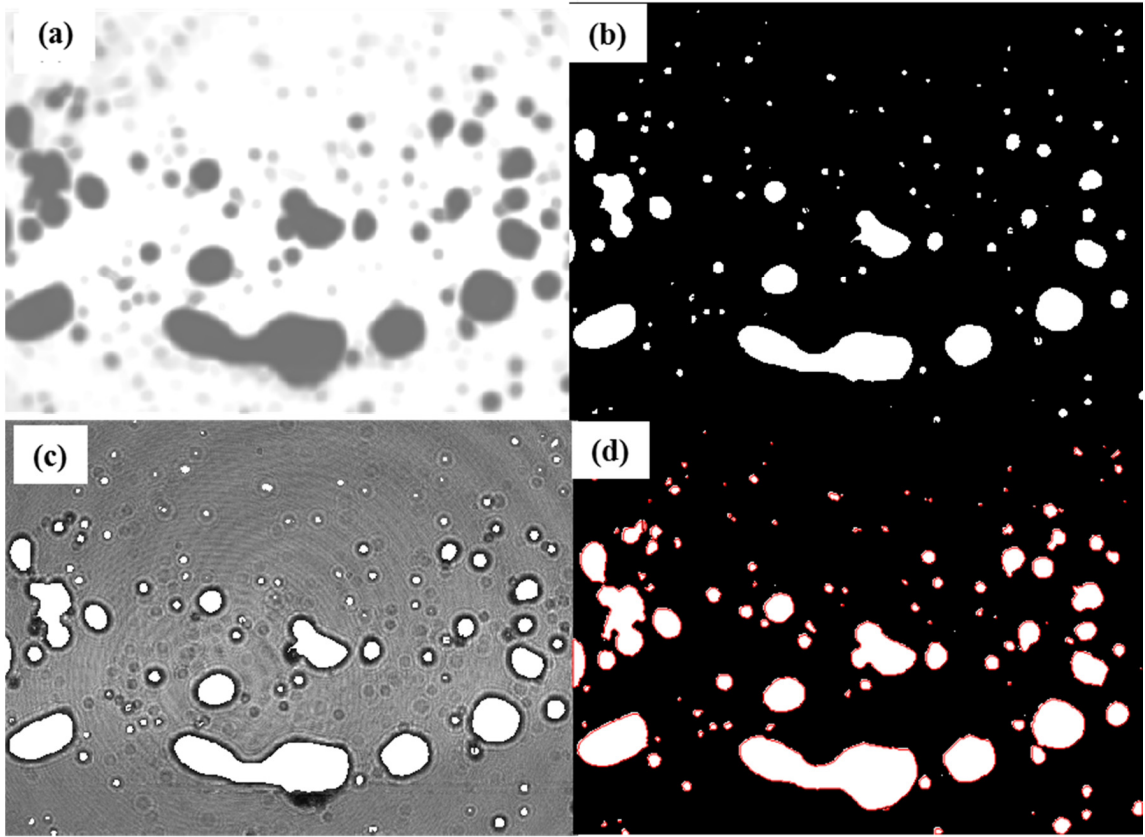

Fig.9. (a) The assistant image and (b) the corresponding marker image. (c) The reconstructed bubble image superimposed with marker image to label the individual bubbles. (d) The binary image showing the result of marker-controlled watershed segmentation using (b) as the marker image. Note that the boundary of individual bubbles are highlighted using red lines in (d).

From the binarized image, large bubbles/clusters are segmented through a modified marker-controlled watershed segmentation algorithm. To obtain makers for watershed segmentation, an assistant image is first generated from the combined reconstruction (Fig. 5a) using the following equations (Kass et al. 1988):

$$I_A = w_L I_L + w_E I_E + w_T I_T \tag{4}$$

$$I_L = G * I_c \tag{5}$$

$$I_E = -|G * \nabla^2 I_c|^2 \tag{6}$$

$$I_T = \frac{\partial^2 I_L / \partial n_\perp^2}{\partial I_L / \partial n} \qquad (7)$$

Of above equations, $I_A$, shown in Fig. 9a, represents the assistant image generated from the original combined image $I_c$. $I_L$ is a filtered version of $I_c$ using a Gaussian filter $G$. The $I_E$ and $I_T$ refer to the image property map representing edge functional, and corners and termination functional, respectively, and $\nabla^2$ is the second-order image derivatives. The $n_\perp$ and $n$ represent unit vectors perpendicular and along intensity gradient direction, respectively. More specifically, $I_E$ is the edge detection result of $I_c$ with a Gaussian filter to remove the noise in the image, and $I_T$ corresponds to the curvature of the lines in a Gaussian-filtered image (i.e. $I_L$ in our case) which can be used to detect the corners and terminations in an image. Finally, the $w_L$, $w_E$, and $w_T$ are the weights of $I_L$, $I_E$ and $I_T$ terms, respectively. The large $w_L$ and $w_E$ terms with small $w_T$ are preferable in generating good assistant images for the segmentation since this operation should effectively preserve the local pixel intensity minima corresponding to bubble centroids and enhance edges of individual bubbles in clusters while suppressing small variation in pixel intensity caused by noise. The resulted assistant image (Fig. 9a) is then employed to generate a marker image for segmentation through extended H-minima transform following Soille (2003). In the assistant image, the extended H-minima transform detects the connected regions of pixels (i.e. the markers) with intensity lower than their neighboring pixels by a value larger than or equal to a predefined threshold through binarization. As shown in Fig. 9b, the markers extracted from this operation (i.e. the white pixels in the image) situate in the center portion of each bubble. As the last step, the combined image is segmented through marker-controlled watershed segmentation using maker image Fig. 9(b) (Gonzalez et al. 2009). Comparing to the traditional watershed segmentation using the distance transform for the binary images (e.g. Meyer 1994), the proposed marker-controlled method only employs a segmentation line at the midpoint of and perpendicular to the line connected two adjacent marker centroids, which minimizes the over-segmentation issue. As illustrated in Fig. 9(c), the generated markers can correctly label all the large objects, and Fig. 9(d) shows the corresponding segmentation result which clear segmentation of large objects from the background. We acknowledge that the proposed method has under-segmentation issue when dealing with the bubbles closely clustered and sharing a common low-intensity centroid region. However, without prior knowledge for discriminating bubble clusters and irregular shape bubbles, this approach provides the appropriate guess of individual bubbles within the clusters. The assessment of the algorithm in the next section through synthetic bubble holograms will provide further demonstration on the accuracy of the segmentation of the bubble clusters using the current method.

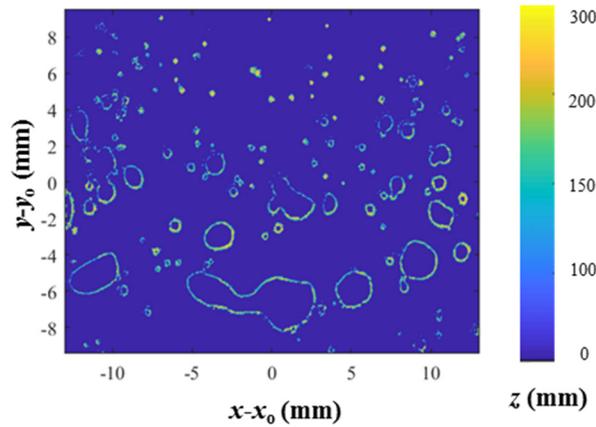

Fig.10. The map of the depth of each pixel along the edges of the segmented bubbles from Fig. 9(d). The coordinate $(x_0, y_0)$ refers to the center of the image and the magnitude of $z$ is the distance between the in-focused location of each pixel and the DIH image plane.

The longitudinal location (i.e. the depth) of each segmented large bubble is calculated as the mean depth of all the edge pixels of the bubble. Specifically, the depth of each edge pixels in reconstructed image is determined using a wavelet-based focus metric by the following equation (Pertuz et al. 2013):

$$M(x, y, z) = |HL(x, y, z)| + |LH(x, y, z)| + |HH(x, y, z)| \tag{8}$$

In the above equation, the metric $M(x, y, z)$ is the sum of the $L$1-norm of three filtered images of the 3D reconstructed optical field, referred to as $HL$, $LH$ and $HH$. The $HL(x, y, z)$ is obtained by applying high pass Harr wavelet to the horizontal (i.e. $x$) direction and low pass Harr wavelet to the vertical (i.e. $y$) direction of the image. Using the same wavelets, the $LH(x, y, z)$ is calculated through the application of low pass to horizontal and high pass wavelet filter to the vertical direction. Similarly, $HH(x, y, z)$ is obtained by applying high pass filter to both the horizontal and the vertical directions. These three wavelet-filtered versions are used to accentuate the horizontal (i.e. $HL$), the vertical (i.e. $LH$) and the diagonal (i.e., $HH$) variation of the pixel intensity of the reconstructed image. In addition, the fourth wavelet-filtered version (i.e.,$LL$), obtained by applying low pass filter to both horizontal and vertical directions, is usually used to remove the noise in the original image. In this paper, the $LH$, $HL$ and $HH$ for calculating the focus metric $M(x, y, z)$ are generated from the $LL$ version of the original 3D reconstructed optical field. Subsequently, the depth of each pixel is determined as the location of the prominent peak in the longitudinal profile of the focus metric using the same approach for extracting the longitudinal locations of small bubbles discussed earlier. Note that the pixels with no prominent peak in their longitudinal focus metric profile are excluded in the calculation of bubble depth. Additionally, a 1×3 pixel moving average filter is applied to the depth values of edge pixels to reduce the effect of random fluctuations in the calculation. As shown in Fig. 10, the edge detection results of the segmented bubbles are superimposed onto the depth map of individual pixels for measuring the longitudinal location of individual large bubbles/clusters. Comparing to previous method that determines the degree of focus through intensity gradient (Guildenbecher et al. 2013, and Gao et al. 2014), the focus metric approach employed in the current study eliminates the need for manually choosing window of specific size to calculate intensity gradient using Sobel operator. In comparison with other wavelet-based focus metric approaches (e.g., Wu et al. 2014), the Haar wavelet adopted in our study uses single-pixel support window which enables more accurate characterization of the intensity gradient across the bubble edge, particularly for small bubbles. In addition, according to Weeks (2006), using the wavelet groups with large support window is susceptible to the influence of virtual image noises which can lower the accuracy of the depth measurement.

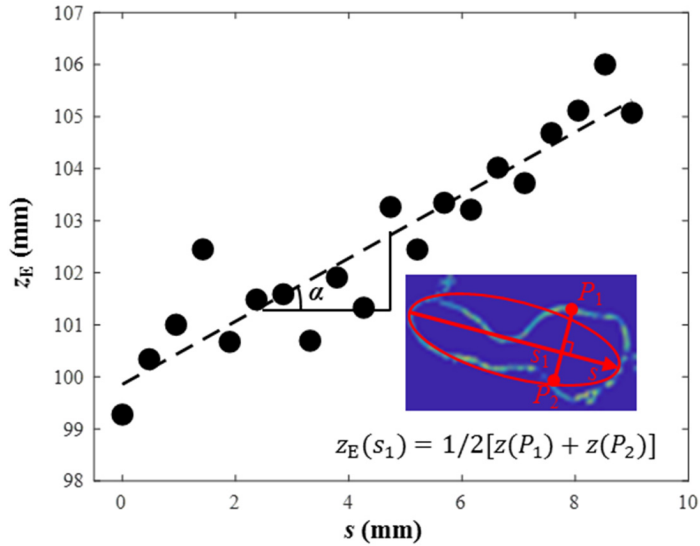

Fig.11. The measurement of the inclination angle of a large bubble sample with respect to the recording plane. The inset figure shows the edge depth map of a bubble sample from Fig. 10(b). Note that the dashed line is the linear regression result of the measurements. For this specific sample bubble, its inclination angle with respect to the recording plane is measured to be +31.2°.

Following the binarization and the bubble depth calculation, the shape and the size of the individual large bubbles/clusters are determined using the method proposed by Karn et al. (2015b). In this analysis, we first calculate the Heywood Circularity Parameter, i.e. $HCF = P/\sqrt{4\pi A}$ (the perimeter $P$ of the bubble to the perimeter of the circle of the same area), for each bubble and cluster in the image. According to Olson (2011), the bubbles having $HCF$ in the range from 0.9 to 1.15 are considered as spherical and their size is estimated using the same formula as that for the small spherical bubbles described earlier. For bubbles with $HCF$ outside the abovementioned range, they are approximated as ellipsoids, and the ellipse fitting developed by Haralick and Shapiro (1992) is applied to their 2D projection to determine their minor and major axes as well as their eccentricity. According to Sotiriadis et al. (2005), the ellipsoidal approximation of the bubble is generated using the revolution of the 2D projection of the bubble about its major axis. However, such approximation can yield high uncertainty in estimating the volume of a bubble which has a large inclination angle with respect to the recording plane (i.e. *x-y* plane). To reduce such uncertainty, we provide an estimate of the inclination angle (with respect to *x-y* plane) of each large bubble/cluster using the depth map of its edge pixels as shown in Fig. 11. Specifically, the depth of the edge pixels corresponding to each location *s* along the major axis of the bubble, i.e. $z_E(s)$, are determined by the average depth of the pixels at the intersections of the line perpendicular to the major axis with the edge (illustrated in the inset figure in Fig. 11). Subsequently, the slope of the linear regression line derived from $z_E(s)$ vs *s* plot is used to estimate the bubble inclination angle *α* with respect to the *s*. Then the 3D major axis of an ellipsoidal bubble is approximated to be $|s|/\cos\alpha$.

## 4. Assessment of the proposed approach

The assessment of the proposed algorithm is conducted using both the synthetic bubble holograms and a fabricated physical target of pillars with different inclination angles. The former is used to determine the accuracy of the bubble size and location measurement, while the latter is employed for assessing the measurement of inclination angles using the depth variation of edge pixels.

### 4.1. Assessment through synthetic bubble holograms

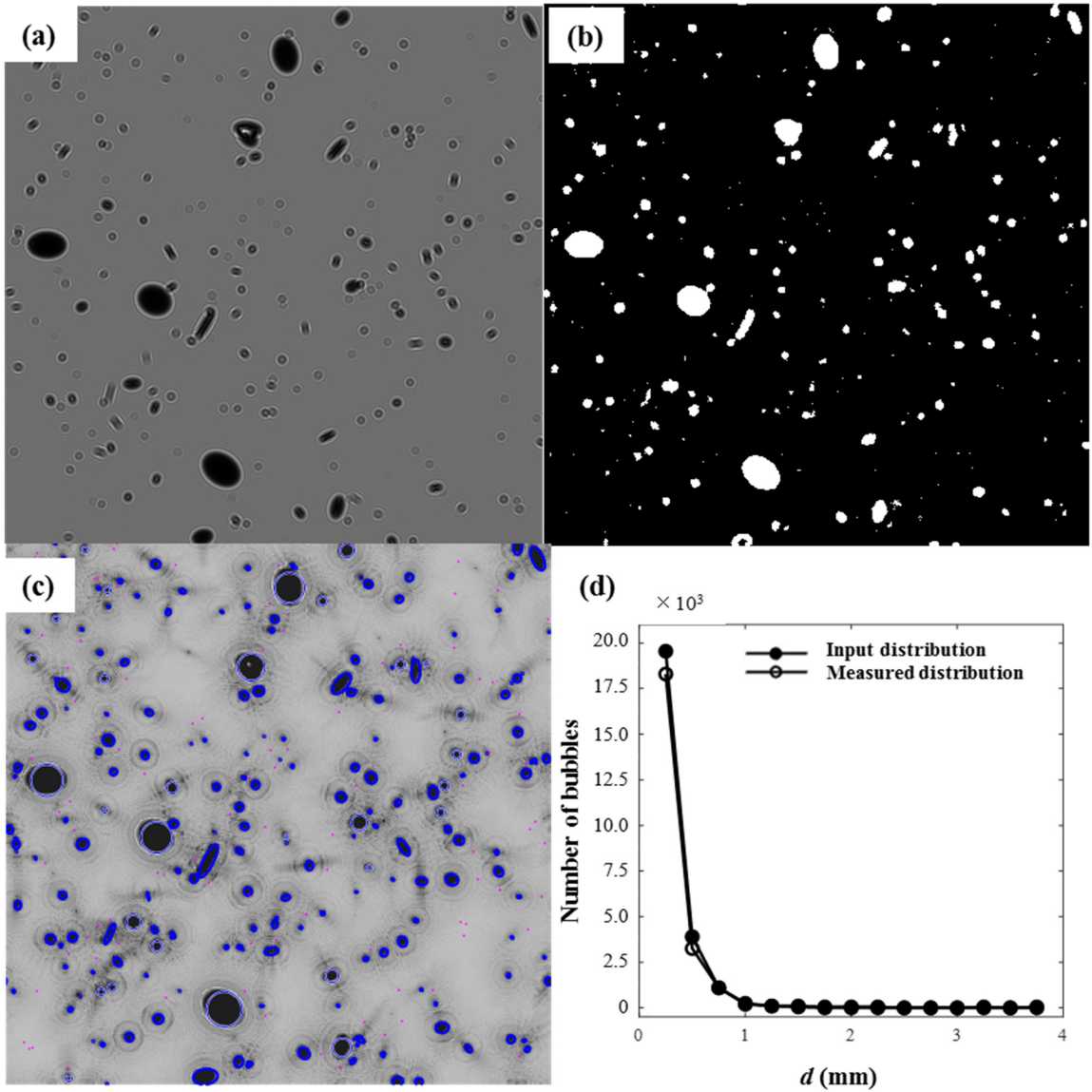

Fig.12. (a) A sample of the synthetic bubble hologram and (b) the corresponding binarized bubble image obtained from the proposed hologram processing approach. (c) The combined image with the edge of small bubbles marked in red and in blue for large bubbles/clusters. (d) A comparison of the ground truth and the measured distribution from the proposed algorithm, where $d$ is the area-equivalent diameter of the synthesized bubbles.

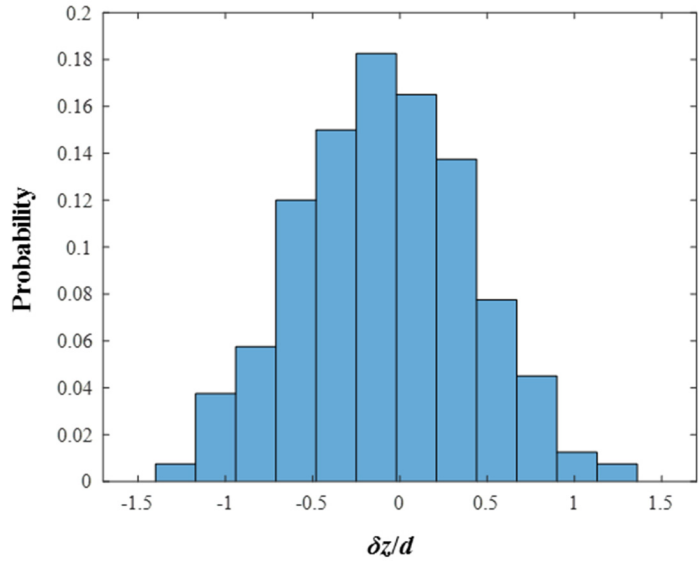

Fig.13. The distribution of the longitudinal positioning error scaled equivalent bubble diameter.

We first assess the measurement of bubble size distribution and location using synthetic bubble holograms. The generation of synthetic holograms is conducted using Rayleigh-Sommerfield diffraction formula following the approach described in the literature (e.g. Zhang et al. 2006, Gao 2014, and Toloui and Hong 2015). To match the hologram settings used in the water tunnel experiments described in Section 2, the synthetic hologram is generated using 51 μm/pixel resolution under 532 nm laser illumination, with bubble size ranging from 0.1 mm to 3.8 mm following log-normal distribution, bubble eccentricity ranging from 0 to 0.85, and the $z$ location from 40 mm to 240 mm ($z = 0$ is at the in-focused plane of DIH measurement). In total 50 synthetic holograms with each consisting of 500 bubbles are used for the analysis. Fig. 12 shows a sample synthetic hologram (Fig. 12a) with the corresponding binarized image (Fig. 12b) and the combined minimum intensity image with the edge of bubbles marked in red for small bubbles and in blue for large bubbles/clusters (Fig. 12c). As shown Fig. 12(d), the resulted bubble size distribution agrees well with the ground truth over the entire size range, though a small difference is observed for bubbles less than 0.2 mm-in-diameter. We attribute such discrepancy to the overlapping of small bubbles with large bubbles in the synthesized bubble holograms. As for the bubble depth measurement, the error $\delta z$ is less than their equivalent diameter for over 90% of bubbles as shown in Fig. 13.

**4.2. Assessment of inclination angle measurement using a physical target**

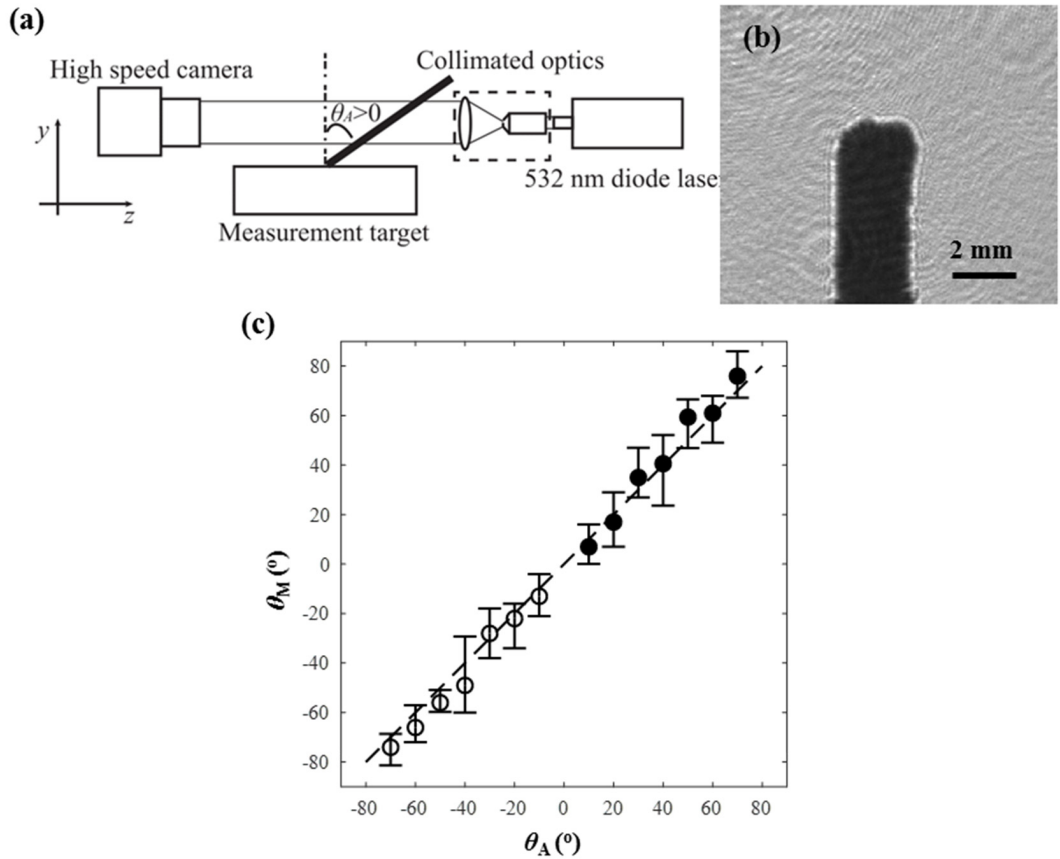

Fig.14. Assessment of inclination angle measurement with the proposed approach. (a) Schematic of the experimental setup. (b) A sample hologram of a pillar with 60° inclination angle with respect to the hologram recording plane. (c) The comparison between the measured inclination angle ($\theta_M$) and the actual angle ($\theta_A$).

The uncertainty for the measurement of inclination angle is assessed using a 3D printed physical target. The target consists a series of pillars of 2 mm-in-diameter and 15 mm-in-length. Their inclination angle with respect to the hologram recording plane changes from -70° to 70° with 10° increment. As shown in Fig. 14(a), the target is placed in the sample volume of the DIH setup operated under the same settings as those for water tunnel experiments. Fig. 14(b) shows a sample hologram from the measurement of a pillar with 60° inclination angle. The inclination angle of each pillar is estimated following the approach described in Section 3 and the uncertainty of each measurement is evaluated using the 90% confidence interval derived from the slope fitting errors from linear regression. As shown in the Fig 14(c), the measured inclination angles ($\theta_M$) match their actual values ($\theta_A$) with an uncertainty about 15 %.

## 5. Measurement of instantaneous gas leakage rate

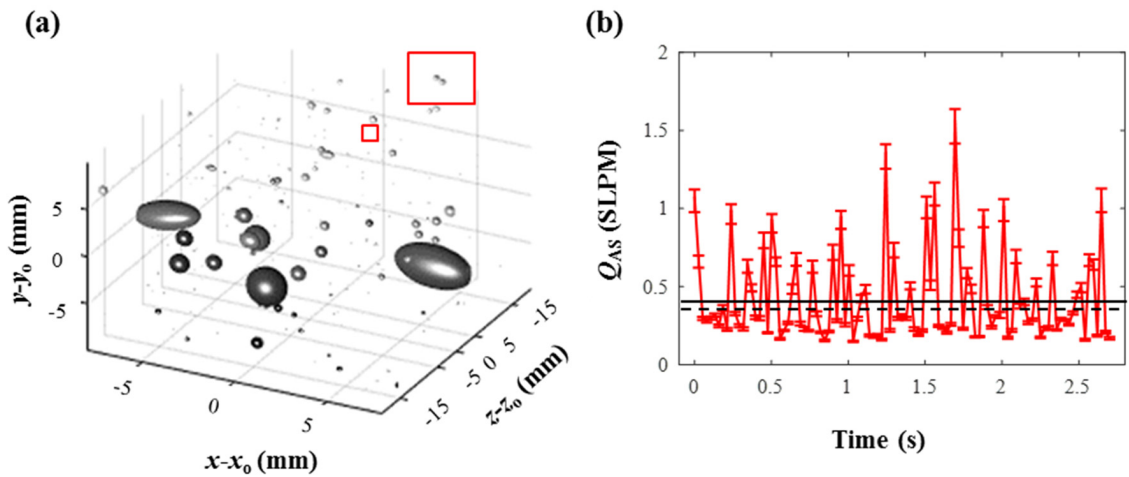

Fig.15. (a) 3D visualization of the bubble reconstruction results from a sample hologram. The coordinate ($x_0$, $y_0$) refers to the center of the image and $z_0$ corresponds to the center location of the tunnel. Inset is a magnified view of the red rectangle shown on the plot to depict the small spherical bubbles captured in the flow. (b) The instantaneous gas leakage rate calculated from the bubble holograms. The solid black line is corresponding to the ventilated rate controlled by the mass flow controller (0.40 SLPM) and the dashed black line is the averaged DIH measurement results of the whole sequence (0.38 SLPM).

The proposed approach is employed to measure the instantaneous gas leakage rate from a ventilated supercavity in the water tunnel experiment described in Section 2. The dataset consists of 8200 holograms captured at 3000 frames/s over a time duration of 2.7 s. Fig. 15(a) shows the 3D visualization of the reconstructed bubble distribution from a sample hologram. The results demonstrate the capability of the proposed approach in capturing the distribution of the bubbles over a wide range of size and the inclination of large bubbles/clusters with respect to the hologram recording plane. Following the approach from Karn et al. (2015b), the instantaneous gas leakage rate can be determined as the gas flux estimated using the gas volume and the velocity of each bubble extracted from the holograms. Note that since the time period for the shed bubbles from the cavity to reach the measurement region is very short (about 0.03 s in our case), the change of gas mass in the bubble due to water absorption during this period is neglected in the estimate of the gas leakage according to Karn et al. (2015b). As shown in Fig. 15(b), the instantaneous air leakage from the ventilated supercavity shows a strong intermittent behavior which requires further investigations. Nevertheless, the average gas leakage rate derived from our measurement over the entire time span is within 5 % from our ventilation input, providing additional assessment on the accuracy and robustness of our technique.

## 6. Conclusions

The paper presents a hybrid bubble hologram processing approach for measuring the size and 3D distribution of bubbles over a wide range of size and shape. The proposed method consists of five major steps, including image enhancement, digital reconstruction, small bubble segmentation, large bubble/cluster segmentation, and post-processing. Specifically, a small bubble exhibits a prominent intensity minimum in its longitudinal intensity profile, which is used to segment it from the 3D reconstructed optical field, and its longitudinal location is determined by the location of its minimum intensity. After excluding the small bubbles from the hologram, the large bubbles/clusters are segmented using a modified watershed segmentation algorithm. Unlike that for the small bubbles, the depth of large bubbles is determined through a wavelet-based focus metric. Furthermore, the inclination angle of large bubbles with respect to the hologram recording plane is determined based on the depth variation along the edge of the bubble on the recording plane. All the above information is employed to obtain accurate estimate of the gas volume contained in the bubble flow. The assessment using the synthetic bubble holograms as well as a 3D printed physical target further demonstrates the capability of our processing code to capture bubbles over a wide range of size and shapes as well as their 3D positions, and also the ability to measure bubble inclination with respect to the recording plane (about 15% uncertainty). Finally, we implement our technique to estimate the instantaneous gas leakage rate from a ventilated supercavity generated in a water tunnel experiment. The measurement showcases the strong intermittent behavior of gas leakage from the cavity, and the measured average gas leakage rate matching within 5% of the ventilation input. Overall, our paper introduces a low cost, compact and high resolution bubble measurement technique that can be used for characterizing low void fraction bubbly flow in a broad range of applications.

## Acknowledgements

This work is supported by the Office of Naval Research (Program Manager, Dr. Thomas Fu) under grant No. N000141612755 and the start-up funding received by Prof. Jiarong Hong from University of Minnesota. We would like to thank the help from Mr. Santosh Kumar for assisting the ventilated supercavitation bubbly wake experiments. The authors also gratefully acknowledge the discussion of the algorithm with Mr. Kevin Mallery.

## References

Bonifazi, G., Serranti, S., Volpe, F., Zuco, R., 1999. A combined morphological and color based approach to characterize flotation froth bubbles. Proc. Second Int. Conf. Intell. Process. Manuf. Mater. 1, 465–470.

Brücker, C., 2000. PIV in two-phase flows. von Karman Institute for fluid dynamics, Lecture Series, 1.

Da Silva, M. J., Schleicher, E., Hampel, U., 2007. Capacitance wire-mesh sensor for fast measurement of phase fraction distributions. Meas. Sci. and Technol. 18(7), 2245.

Estevadeordal, J., Goss, L., PIV with LED: particle shadow velocimetry (PSV) technique, in: 43rd AIAA aerospace sciences meeting and exhibit, 2005.

Felder, S., Chanson, H., 2015. Phase-detection probe measurements in high-velocity free-surface flows including a discussion of key sampling parameters. Exp. Therm. Fluid Sci. 61, 66-78.

Gao, J., 2014. Development and applications of digital holography to particle field measurement and in vivo biological imaging. PhD diss., Purdue University.

Gao, J., Guildenbecher, D.R., Engvall, L., Reu, P.L., Chen, J., 2014. Refinement of particle detection by the hybrid method in digital in-line holography. Appl. Opt. 53(27), 130-138.

Gonzalez, R.C., Woods, R.E., Eddins, S.L., 2009. Digital Image Processing Using MATLAB®. Gatesmark Publishing.

Grubbs, F.E., 1969. Procedures for detecting outlying observations in samples. Technometrics, 11(1), pp.1-21.

Guildenbecher, D.R., Gao, J., Reu, P.L., Chen, J., 2013. Digital holography simulations and experiments to quantify the accuracy of 3D particle location and 2D sizing using a proposed hybrid method. Appl. Opt. 52(16), 3790-3801.

Haralick, R.M., Shapiro, L.G., 1992. Computer and Robot Vision. Addison-Wesley, New York.

Hernandez-Alvarado, F., Kleinbart, S., Kalaga, D.V., Banerjee, S., Joshi, J.B., Kawaji, M., 2018. Comparison of void fraction measurements using different techniques in two-phase flow bubble column reactors. Int. J. Multiph. Flow. 102, 119-129.

Honkanen, M., Eloranta, H., Saarenrinne, P., 2010. Digital imaging measurement of dense multiphase flows in industrial processes. Flow Meas. Instrum. 21 (1), 25–32.

Junker, B., 2006. Measurement of bubble and pellet size distributions: past and current image analysis technology. Bioproc. Biosyst. Eng. 29(3), 185-206.

Kamp, A., Chesters, A., Colin, C., Fabre, J., 2001. Bubble coalescence in turbulent flows: a mechanistic model for turbulence-induced coalescence applied to microgravity bubbly pipe flow. Int. J. Multiph. Flow 27(8), 1363-1396.

Karn, A., Arndt, R.E., Hong, J., 2016. An experimental investigation into supercavity closure mechanisms. J. Fluid Mech. 789, 259-284.

Karn, A., Ellis, C., Arndt, R. E., Hong, J., 2015. An integrative image measurement technique for dense bubbly flows with a wide size distribution. Chem. Eng. Sci. 122, 240-249.

Karn, A., Monson, G.M., Ellis, C.R., Hong, J., Arndt, R.E., Gulliver, J.S., 2015. Mass transfer studies across ventilated hydrofoils: A step towards hydroturbine aeration. Int. J. Heat Mass Transf. 87, 512-520.

Kass, M., Witkin, A., Terzopoulos, D., 1988. Snakes: Active contour models. Int. J. Comput. Vis. 1(4), 321-331.

Katz, J., Sheng, J., 2010. Applications of holography in fluid mechanics and particle dynamics. Annu. Rev. Fluid Mech. 42, 531-555.

Khanam, T., Rahman, M.N., Rajendran, A., Kariwala, V., Asundi, A.K., 2011. Accurate size measurement of needle-shaped particles using digital holography. Chem. Eng. Sci. 66(12), 2699-2706.

Laakkonen, M., Moilanen, P., Miettinen, T., Saari, K., Honkanen, M., Saarenrinne, P., Aittamaa, J., 2005. Local bubble size distributions in agitated vessel: comparison of three experimental techniques. Chem. Eng. Res. Des. 83(1), 50-58.

Lau, Y.M., Deen, N.G., Kuipers, J.A.M., 2013. Development of an image measurement technique for size distribution in dense bubbly flows. Chem. Eng. Sci. 94, 20-29.

Liao, Y., Lucas, D., 2009. A literature review of theoretical models for drop and bubble breakup in turbulent dispersions. Chem. Eng. Sci. 64(15), 3389-3406.

Liao, Y., Lucas, D., 2010. A literature review on mechanisms and models for the coalescence process of fluid particles. Chem. Eng. Sci. 65(10), 2851-2864.

Lin, B., Recke, B., Knudsen, J.K., Jørgensen, S.B., 2008. Bubble size estimation for flotation processes. Miner. Eng. 21 (7), 539–548.

Liu, H., Yu, J., Wang, T., Yang, Y., Wang, J., Zheng, R., 2013. Digital holography experiment of 3D detection of underwater bubble fields. Chin. Opt. Lett. 11(s2), S20901.

Liu, T.J., Bankoff, S.G., 1993. Structure of air-water bubbly flow in a vertical pipe—II. Void fraction, bubble velocity and bubble size distribution. Int. J. Heat Mass Transf. 36(4), 1061-1072.

Meyer, F., 1994. Topographic distance and watershed lines. Signal Process 38 (1), 113–125.

Olson, E., 2011. Particle shape factors and their use in image analysis part 1: theory. J. GXP Compliance. 15(3), p.85.

Otsu, N., 1979. A threshold selection method from gray-level histograms. IEEE Trans. Syst. Man Cybern. SMC-9, 62–66.

Pertuz, S., Puig, D., Garcia, M.A., 2013. Analysis of focus measure operators for shape-from-focus. Pattern Recognit. 46(5), 1415-1432.

Roesler, T., Lefebvre, A., 1989. Studies on aerated-liquid atomization. Int. J. Turbo. Jet-Engines 6(3-4), 221-230.

Saberi, S., Shakourzadeh, K., Bastoul, D., Militzer, J., 1995. Bubble size and velocity measurement in gas—liquid systems: Application of fiber optic technique to pilot plant scale. Can. J. Chem. Eng. 73(2), 253-257.

Sahoo, P., Wilkins, C., Yeager, J., 1997. Threshold selection using Renyi's entropy. Pattern Recognit. 30 (1), 71–84.

Sahoo, P.K., Arora, G., 2004. A thresholding method based on two-dimensional Renyi's entropy. Pattern Recognit. 37 (6), 1149–1161.

Sentis, M.P., Onofri, F.R., Lamadie, F., 2018. Bubbles, drops, and solid particles recognition from real or virtual photonic jets reconstructed by digital in-line holography. Opt. Lett. 43(12), 2945-2948.

Shao, S., Wu, Y., Haynes, J., Arndt, R.E., Hong, J., 2018. Investigation into the behaviors of ventilated supercavities in unsteady flow. Phys. Fluids, 30(5), 052102.

Smith, J.S., Burns, L.F., Valsaraj, K.T., Thibodeaux, L.J., 1996. Bubble column reactors for wastewater treatment. 2. The effect of sparger design on sublation column hydrodynamics in the homogeneous flow regime. Ind. Eng. Chem. Res. 35(5), 1700-1710.

Soille, P., 2003. Morphological Image Analysis: Principles and Applications, 2nd ed. Springer-Verlag, New York.

Sotiriadis, A.A., Thorpe, R.B., Smith, J.M., 2005. Bubble size and mass transfer characteristics of sparged downwards two-phase flow. Chem. Eng. Sci. 60(22), 5917-5929.

Talapatra, S., Sullivan, J., Katz, J., Twardowski, M., Czerski, H., Donaghay, P., Hong, J., Rines, J., McFarland, M., Nayak, A.R., Zhang, C., Application of in-situ digital holography in the study of particles, organisms and bubbles within their natural environment, in: Ocean Sensing and Monitoring IV, 2012.

Tian, L., Loomis, N., Domínguez-Caballero, J. A., Barbastathis, G., 2010. Quantitative measurement of size and three-dimensional position of fast-moving bubbles in air-water mixture flows using digital holography. Appl. Opt. 49(9), 1549-1554.

Toloui, M. and Hong, J., 2015. High fidelity digital inline holographic method for 3D flow measurements. Opt. Express. 23(21), 27159-27173.

Weeks, M., 2010. Digital signal processing using MATLAB & wavelets. Jones & Bartlett Learning.

Wosnik, M., Arndt, R.E., 2013. Measurements in high void-fraction bubbly wakes created by ventilated supercavitation. J. Fluids Eng. 135(1), 011304.

Yingchun, W., Xuecheng, W., Jing, Y., Zhihua, W., Xiang, G., Binwu, Z., Linghong, C., Kunzan, Q., Gréhan, G., Kefa, C., 2014. Wavelet-based depth-of-field extension, accurate autofocusing, and particle pairing for digital inline particle holography. Appl. Opt. 53(4), 556-564.

Zhang, Y., Shen, G., Schröder, A., Kompenhans, J., 2006. Influence of some recording parameters on digital holographic particle image velocimetry. Opt. Eng. 45(7), 075801.